\documentclass[preprint,prb]{revtex4}
\usepackage{natbib}
\bibpunct{}{}{,}{s}{,}{,}

\usepackage{graphicx}
\usepackage{epsfig}
\usepackage{ae}
\usepackage{aecompl}

\begin{document}
\bibliographystyle{apsrev}

\title{Self-Consistent-Field Study of the Alignment by an Electric Field
of a Cylindrical Phase of Block Copolymer}

\author{Chin-Yet Lin and M. Schick}
\affiliation{Department of Physics, University of Washington,  Box
  351560, Seattle, WA 98195-1560 U. S. A.}

\date{April 10, 2006}

\begin{abstract} Self-Consistent Field Theory is applied to
a film of cylindrical-forming block copolymer subject to a surface
field
which tends to align the cylinders parallel to electrical plates, and to
an external electric field tending to align them
perpendicular to the plates. The Maxwell
equations and self-consistent field equations are solved exactly,
numerically, in real space. By comparing the free energies of different
configurations, we show that for weak surface fields, the phase of cylinders
parallel to the plates makes a direct transition to a phase in which
the cylinders are aligned with the field throughout the sample.
For stronger surface fields, there is an
intermediate phase in which cylinders in the interior of the film,
aligned with the field, terminate near the plates.  For surface fields
which favor the minority block, there is a boundary
layer of hexagonal symmetry at the plates
in which the monomers favored by the surface field
occupy a larger area than they would if the cylinders extended
to the surface.
\end{abstract}

\maketitle

\newpage \section {INTRODUCTION}

With their ability to self-assemble, block copolymers are a natural choice
of material to be utilized in the fabrication of devices incorporating
periodic arrays \citep{morkved96,park97,walheim99,talbrecht00}. A major
difficulty with the process, however, is that the system rarely forms a
single domain of such an array, so a desirable long-range order is absent.
Furthermore, preferential interactions between the blocks and substrate
can cause the array to be aligned in a direction other than the one
desired for use. Alignment of domains can be obtained in several ways, of
which the one of interest to us is the use of an electric field
\citep{talbrecht00,talbrechtsci00}. This takes advantage of the fact that
the two blocks in general have different dielectric constants, so that it
is favorable for the array to align itself such that there is as little
induced polarization charge as possible\citep{amundson94}. For lamellar-
and cylinder-forming diblocks, this means that a sufficiently strong field
will align the array with lamellae or cylinders oriented with their long
symmetry axis parallel to the applied field, perpendicular to the
electric plates.

If one assumes that the system, in zero external electric field, is
characterized by layers of cylinders lying parallel to the substrate and,
in large external fields, is characterized by all cylinders oriented
perpendicular to the field, then one would like to know both the equilibrium
morphology at intermediate field strengths and the phase diagram of the
system. Two possibilities were suggested by Thurn-Albrecht et al.
\citep{talbrecht00}. In the first, the electric field has essentially
little effect on the arrangement of parallel cylinders  up to the field strength
at which the perpendicular morphology becomes the globally stable phase.
If this situation obtains, one expects that the difference in
electrostatic energies of the two configurations would be $E_{c}^2{\cal
A}d_0$, with ${\cal A}$ the area of each plate, and $d_0$ the distance
between plates. Hence a first-order transition would take place at a
critical electric field, $E_{c},$ such that $E_{c}\propto d_0^{-1/2}.$ In
the second scenario, an intermediate phase is formed, one in which
the surface field is sufficiently strong that a boundary layer near the
plates has a different morphology than in the rest of the film.
As Thurn-Albrecht et al.
reported observing cylinders both in parallel and perpendicular
orientations {\em over a range of field strengths}, the implication was that
while the cylinders were oriented perpendicular
to the substrate in most of the film,
the boundary
layer might consist of cylinders oriented parallel to the
surface.
The difference in
electrostatic energies between the perpendicular phase and this
intermediate phase is, then, not $E_c ^2{\cal
A}d_0,$ but $E_c ^2{\cal A}t$, where $t$ is a length characteristic of the
thickness of the boundary layer. Presumably it depends
only weakly on the film thickness, $d_0$. Thus
one expects that $E_{c}$ also varies only weakly with thickness
for thick
films.  Thurn-Albrecht et al.\cite{talbrecht00} did indeed find that
the critical electric field, beyond which only the perpendicular phase
was observed, was relatively constant for large film thicknesses.

The existence of an intermediate phase is a consequence of the
competition between the surface fields, which prefer a phase in which
the cylinders are parallel to the surface, and the electric field which
prefers that the cylinders be aligned with it, perpendicular to the
substrate. If the system were semi-infinite, the intermediate phase
would simply be labeled a surface phase, one of different symmetry than
that of the bulk, a very common situation \cite{arnold02}. Such a phase
would be expected to persist in the case of a finite, thick
film. However when the film is sufficiently thin, the clear distinction
between bulk and surface properties can no longer be made, and the two
orientations compete throughout the film, resulting in the elimination
of one of them.

Most theoretical work on block-copolymers in electric fields has focused
on lamellar-forming phases, as opposed to cylindrical ones, and on the
dynamic mechanisms by which morphologies could realign,
rather than on the equilibrium morphologies themselves. Such dynamic
mechanisms were the subject of the seminal papers
of Amundson et al. \cite{amundson93,amundson94}, and of more recent
dynamic density functional
calculations \cite{kyrylyuk02,lyakhova04,lyakhova06}.
The form of the equilibrium phases
was taken up by Pereira and Williams \cite{pereira99} and by Tsori and
Andelman \cite{tsori02}. Both groups considered one intermediate phase in
which a single lamella at the plates remains parallel to them,
although surface fields could conceivably lead to a sequence of
intermediate phases distinguished by the number of lamellae
parallel to the surface. Such a discrete increase, or decrease, of the
number of lamellae with applied field
would be analogous to the layer-by-layer growth modes observed in
adsorbed films \cite{pandit82}. Both groups presented phase
diagrams as functions of applied
electric field and film thickness. Although similar, they differ in one
respect: that of
Tsori and Andelman shows the perpendicular phase to  be favored always at
sufficiently large external fields, which is certainly correct. The
effects of surface fields have also been explored
\cite{tsori02,lyakhova04,lyakhova06}. As expected, a strong surface field favors
the formation of an intermediate phase.
The evolution of morphology with external field
in a single lamella has recently been
studied by Matsen using self-consistent field theory \cite{matsen05}.

As for cylindrical-forming diblocks in an external electric field, the
only published work has again focused on
the mechanisms of morphology realignment. This has been studied
experimentally \cite{xu05}, and also theoretically
by dynamical self-consistent field
theory \cite{xu05,lyakhova06}. The advantage of this technique is that one
follows the dynamics of the system, and one can clearly see how
different are the pathways towards an equilibrium morphology
depending upon the initial configuration.  The actual equilibrium
morphologies, which are the focus of our interest, are difficult to
achieve however, just as they often are in experiment.

Some insight into the expected phase diagram of the cylinder-forming
diblock copolymer
system can be garnered from the behavior of the lamellar-forming
ones. In both systems, surface fields prefer an orientation which is
different from that preferred by the electric field. It should be noted
that this holds irrespective of whether the surface fields prefer one
monomer species, $A$, or the other, $B$. Thus we should expect to obtain
a phase of cylinders parallel to the surface when the electric field is
weak, a phase of perpendicular cylinders when the electric field is
strong, and an intermediate phase for thick films, and ones subject to
large surface fields. The principal difference between the cylinder- and
lamellar-forming systems will be in the symmetry of the surface
phase. Similarly the principle difference between cylinder-formers adsorbed on a
substrate that prefers the minority monomer, $A$, and those adsorbed on
a surface preferring the majority monomer, $B$, will be in the specific morphology
of the surface phase. Otherwise the general form of the phase diagrams
of these systems is expected to be similar due to the identity of the underlying
physics.

In this paper, we study a cylinder-forming block-copolymer system
in external electric fields and examine the morphology and phase
behavior in the presence of surface fields which prefer the minority
monomer. In section II, we first show that
some general features of the surface phase diagram in the electric
field-chemical potential plane are easily discerned from basic
thermodynamics, and without detailed calculation. However thermodynamics
does not tell us the nature of the intermediate phases, while specific
calculation does.  We then turn to self-consistent field theory, and solve
exactly the self-consistent equations and the Maxwell equations in the
manner used by us previously for a bulk phase\citep{lin05}. Subsequently,
the same idea has been employed with a different technical implementation
in other systems \cite{matsen05,matsen06}. As a consequence of solving the
Maxwell equations exactly, we do not assume, as in refs.
\cite{xu05,lyakhova04, lyakhova06}, that the effect of the electric field is weak;
that is, that the fractional variation of the dielectric constant
throughout the morphology is small \cite{tsori06}. Our results are
presented in section III.  For weak surface fields, or thin films,
there is no intermediate phase, and we find the first of
the scenarios suggested by Thurn-Albrecht et al\cite{talbrecht00}.
The phase of parallel cylinders transforms directly to one of perpendicular
cylinders. These cylinders are, of course, affected in their structure by
the presence of the substrate surface fields, but a cross-section parallel
to the substrate confirms the hexagonal order, even very close to the
substrate.

For stronger surface fields or thicker films, we find an intermediate
phase in which a layer adjacent to the plates has a different morphology
than that of the field-aligned cylinders in the rest of the film. The
morphology is not one of cylinders parallel to the plates, but has the
same hexagonal symmetry as a cross-section of the field-aligned aligned
cylinders. However the distribution of monomers is altered so that the
monomer favored by the surface field occupies a larger surface area than
it would had the cylinders extended to the surface.

We conclude with a brief discussion and comparison of our results
with experiment. In particular we find, in agreement with experiment
in the presence of strong surface fields, that the phase boundary
between the perpendicular phase and the intermediate phase
is a weak function of film thickness for large thicknesses.
Along this locus of first order transitions,
the perpendicular and intermediate phases
coexist. However as we just noted, the boundary layer of the
intermediate phase does not consist of cylinders parallel to the
substrate. Thus parallel cylinders do not exist in either
phase, and one would not observe {\em in equilibrium}, along this
particular phase boundary, the coexistence of cylinders in both
parallel and perpendicular orientations.

\section{Theory} \subsection{Thermodynamics} Because we are dealing with a
surface film, we consider the excess free energy per unit area, or surface
free energy per unit area, $ \sigma$, defined as follows.  Let $F_{tot}$
be the total free energy of the system of volume $\Omega=d_0{\cal A}$ such
that the free energy per unit volume,
\begin{equation} f_b\equiv \lim_{\Omega\rightarrow\infty}\frac{F_{tot}}{\Omega},
\end{equation}
is the Legendre transform of the energy per unit volume, $e_b$, with respect to the entropy density, $s_b$, number density, $\rho_b$,  and displacement field  i.e.
\begin{equation}
f_b=e_b- s_bT-\rho_b\mu-{\bf D}_b\cdot{\bf E},
\end{equation}
with ${\bf D}_b$ and ${\bf E}$ the volume-averaged
displacement and electric fields. Then the excess free energy per unit
area, $\sigma$, is
\begin{equation}
\sigma\equiv \lim_{{\cal
A}\rightarrow \infty}\frac{F_{tot}-f_b\Omega}{\cal A}
\label{sigma}
\end{equation}
with the differential
\begin{equation}
d\sigma =-s_sdT-\rho_s d\mu- D_{s}dE_{ext}.
\label{difsigma}
\end{equation}
Here $s_s$ is the excess entropy per unit area, $\rho_s$ the excess number
of particles per unit area,  and $E_{ext}$ is the externally applied field which is equal to the spatially
averaged electric field
\cite{lin05}.
The quantity $D_{s}$ is the magnitude of the
{\em surface contribution} to the displacement field, or surface excess
displacement field
\begin{equation}
{\bf D}_s\equiv\lim_{{\cal
A}\rightarrow \infty}\frac{1}{{\cal A}}\int [{\bf D}({\bf r})-{\bf D}_b({\bf r})]d^3 r,
\end{equation}
where ${\bf D}_b({\bf r})$ is the displacement
field in the bulk cylindrical phase.
The surface excess displacement field is in the $x$
direction, normal to the plates.

From this differential, one easily finds a
Claussius-Clapeyron equation pertaining to the surface phases. If one plots the
phase diagram at fixed temperature in the electric field-chemical potential
plane, then the slope of the boundary between surface phases, $dE_{ext}/d\mu$,
is given by
\begin{equation}
\frac{dE_{ext}}{d\mu}=-\frac{\Delta\rho_s}{\Delta D_s},
\label{cc}
\end{equation}
where $\Delta\rho_s$ and $\Delta D_s$ are the differences in the excess surface
densities and displacement fields in the two phases.  At zero external field,
there are an infinite number of parallel phases distinguished by the
integer number of layers of cylinders, and a perpendicular phase in which
the cylinders are perpendicular to the substrate. The transitions between
all phases are generally first-order; those between the perpendicular and parallel
phases are first-order due to the difference in symmetry between them;
those between different parallel phases are generally first order because an
additional layer of parallel cylinders cannot be added in a continuous
manner.  As one turns on the electric field, the phase space of the perpendicular
phase will increase, and that of the parallel phases will decrease. It is
clear that the spatially averaged polarization will be smaller
in the parallel phases than in the perpendicular phase, and so the
displacement field will decrease discontinuously on entering a parallel
phase from a perpendicular phase, and will increase discontinuously when
leaving a parallel phase. Further, the surface density is presumably
a monotonically
increasing function of the chemical potential. Hence we expect from eq.
\ref{cc} to find the infinite number of parallel phases to be enclosed by
phase boundaries which, if no other phases are encountered, will be
in the shape of wickets. The legs of each wicket meet the
$E_{ext}=0$ axis perpendicularly, because $\Delta D_s=0$ in the absence
of a field. The
top of each wicket is locally horizontal, and the field at which it
occurs defines the critical voltage at which that parallel
phase disappears.
In the region of phase space in which an intermediate phase appears, the
Claussius-Clapeyron equation can be applied to the phase boundaries
between it and the parallel phases, and between it and the perpendicular
phase to obtain useful information relating
the surface densities and displacement fields in the coexisting phases.
It can also be applied, of course, to the lamellar-forming systems
\cite{pereira99,tsori02}.

Another useful result,
\begin{equation}
D_s=- \frac{\partial\sigma}{\partial E_{ext}},
\label{displacement}
\end{equation}
follows from the differential eq. \ref{difsigma}.
We note that, whereas that component of the bulk, spatially averaged,
displacement which is along the direction of the external field
must be positive \cite{landau84},
that same component of the  excess surface
displacement need {\em not} be. In fact, as we expect that the presence of the
surface can only disrupt the perfect alignment of the cylinders with the
electric field, at least near the surfaces, we anticipate that $D_s$
will be {\em negative}. It follows from eq. \ref{displacement} that the
excess free energy per unit area will {\em increase} with increasing electric field.
This contrasts with the behavior of the {\em total} free energy, $F_{tot}$,  and
free energy per unit volume, $f_b$,
which {\em decrease} with increasing external field.

\subsection{Self-Consistent Field Theory}
The method we employ has been described previously \cite{lin05} for a
bulk system, and simply needs known modifications for the case of a
surface film\cite{matsen97}. Thus  its
presentation here can be brief. We consider a melt of $n$ A-B diblock
copolymer chains, each of
polymerization index $N=N_A+N_B$.  The mole fraction of the A-monomers, $f_A=
N_A/N$. We also assume that the Kuhn lengths of the $A$ and $B$
components are identical, a length denoted $a$. The system is confined
between identical plates, a distance $d_0$ apart, each of area ${\cal A}$,
which exert a surface field,
$h({\bf r}),$ preferentially on one block with respect to the other. The
plates are normal to the $x$ axis, and the applied field is along the
 positive direction. The fraction of the volume occupied by
polymer is denoted $\Phi_0({\bf r}),$ and is unity
in the film except near the plates where it falls to zero.

Self-consistent field theory leads to a free energy,  {\cal F},
which is a functional of unknown fields, $W_A$, $W_B$, $\Xi$,
and the unknown electric potential, $V({\bf r})$, and a function of
temperature, $T$, volume, $\Omega$, and area, ${\cal A}$. In an
ensemble in which the external electric potential, $V_0$, is held fixed
\cite{landau84}, the free energy can be written
\begin{eqnarray}
\label{intfreeenergy} \frac{{\cal F}(W_A,W_B,\Xi,V;T,\Omega,{\cal A})}{nk_B
T} &=& -\ln\mathcal{Q}[W_A,W_B]\nonumber \\
&+&\frac{1}{\Omega}\int d\mathbf{r} \{ \chi N\Phi_A\Phi_B -
W_A\Phi_A - W_B\Phi_B\nonumber \\
 &-& h({\bf r})N(\Phi_A-\Phi_B) -\Xi(\Phi_0-\Phi_A-\Phi_B)\nonumber \\
&-&\frac{\Omega\epsilon_0\kappa({\bf r})}{2nk_BT}(\nabla V({\bf r}))^2\}  ~,
\end{eqnarray}
where $\epsilon_0$ is the vacuum permitivity, $k_B$ is Boltzmann's
constant, and $\Phi_A({\bf r})$ and
$\Phi_B({\bf r})$ are the local volume fractions of A and B
monomers.  The function $\mathcal{Q}[W_A,W_B]$ is the partition
function of a single polymer chain
subject to the fields $W_A({\bf r})$ and $W_B({\bf r})$, as is
given below. The field  $\Xi({\bf r})$ is a Lagrange multiplier
that enforces locally the incompressibility constraint,
$\Phi_A({\bf r})+\Phi_B({\bf r})=\Phi_0$.
The dependence on temperature, $T$, comes from the usual Flory
interaction parameter, $\chi$, which to a good approximation is
inversely proportional to $T$.
Finally the local dielectric constant is
$\kappa({\bf r})$. A constitutive relation between this
local dielectric constant and the local volume fractions, $\Phi_A({\bf r})$ and
$\Phi_B({\bf r})$, must be specified. We have chosen, as previously
\cite{lin05},
that the local dielectric constant be given by its local
average. Here
that choice is expressed
\begin{equation}
\kappa({\bf r})=\Phi_0[\kappa_A\Phi_A({\bf r})+\kappa_B\Phi_B({\bf r})]+1-\Phi_0,
\label{constitutive}
\end{equation}
where $\kappa_A$ and $\kappa_B$ are the dielectric constants of the pure
$A$ and $B$ homopolymer phases respectively. Note that near the plates,
the dielectric constant approaches unity. We stress that, while this
choice is clearly correct in the limiting cases of the pure systems and in
the weak segregation limit, it is simply a {\em choice}. While there are
theories of the
constitutive relation appropriate for dilute gases and for critical
fluids, little is known of the detailed relation for dense, solid,
multi-component systems.
Hence that of eq. \ref{constitutive}, which  has the virtue of being both
reasonable and simple, has been commonly employed
\cite{lyakhova04,lin05,xu05,matsen05,matsen06,lyakhova06}.

The requirement that the free energy functional be an extremum with
respect to variation of the electric potential, $V({\bf r}),$
$W_A({\bf r})$, $W_B({\bf r})$, $\Xi({\bf r})$, and of the volume
fractions $\Phi_A({\bf r})$ and
$\Phi_B({\bf r})$ at constant temperature, or $\chi N$,
leads to the following set of equations:
\begin{eqnarray}
\label{scf0}
0&=&\nabla\cdot[\epsilon_0\kappa({\bf r})\nabla  V({\bf r})]~,\\
\label{sceq1}
w_A({\bf r})&=&\chi
N\phi_B({\bf r})+\xi({\bf r})-h({\bf r})N-\frac{\Omega\epsilon_0}{2nk_BT} \Phi_0({\bf r})\kappa_A
(\nabla V({\bf r}))^2 ~,\\
w_B({\bf r})&=&\chi
N\phi_A({\bf r})+\xi({\bf r})+h({\bf r})N -\frac{\Omega\epsilon_0}{2nk_BT}\Phi_0({\bf r})
\kappa_B(\nabla V({\bf r}))^2 ~,\\
\Phi_0({\bf r})&=&\phi_A({\bf r})+\phi_B({\bf r}) ~,\\
\phi_A({\bf r})&=&-\frac{\Omega}{\mathcal{Q}}
\frac{\delta\mathcal{Q}}{\delta w_A({\bf r})} ~,\\
\label{sceq5}
\phi_B({\bf r})&=&-\frac{\Omega}{\mathcal{Q}}\frac{\delta\mathcal{Q}}{\delta
w_B({\bf r})} ~.
\end{eqnarray}
The functions $W_A$, $W_B$, $\Xi$, $\Phi_A$ and $\Phi_B$, which
satisfy these equations are denoted by lower case letters, $w_A$,
$w_B$, $\xi$, $\phi_A$, $\phi_B$, respectively. The first of these
equations is simply Gauss's law for a system with no free charge. The free
energy within the self-consistent field approximation, $F_{\rm scf},$ is obtained by
substitution of these functions into the free energy of
eq.~(\ref{intfreeenergy}),
\begin{equation}
\label{mffreeenergy} F_{\rm scf}(V_0,T,\Omega,{\cal A})={\cal
F}(w_A,w_B,\xi,V;T,\Omega,{\cal A}) ~,
\end {equation}
or
\begin{equation}
\frac{F_{\rm scf}}{nk_B T} = -\ln\mathcal{Q}[w_A,w_B]
-\frac{1}{\Omega}\int d\mathbf{r}\{ \chi
N\phi_A(\mathbf{r})\phi_B(\mathbf{r}) +\xi(\mathbf{r})\Phi_0({\bf r})
+\frac{\Omega\epsilon_0}{2nk_BT}[1-\Phi_0({\bf r})][\nabla V({\bf r})]^2\}
~.\label{mffreeenergy1}
\end{equation}

The nature of the particular system being described by self-consistent
field theory
is specified by the partition function of a representative member of the
entities which comprise the system, in this case, block copolymers.  Therefore
${\mathcal
Q}[w_A,w_B]=\int d{\bf r}\ q({\bf r},1)/k$, where $q({\bf r},s)$
satisfies the modified diffusion equation
\begin{equation}
\frac{\partial q}{\partial s} = \frac{1}{6}Na^2\nabla^2q-w_A({\bf
r})q,\qquad\ {\rm if~~}\ 0\leq s \leq f_A ~,
\end{equation}
and
\begin{equation}
\frac{\partial q}{\partial s} =
\frac{1}{6}Na^2\nabla^2q-w_B({\bf r})q, \qquad \ {\rm if~~}\ f_A\ <\ s\
\leq 1 ~,
\end{equation}
with the initial condition $q({\bf r},0)=1,$ and $k$ is a volume of no
consequence here.

The total density profile, $\Phi_0({\bf r})$, and the surface field,
$h({\bf r})$, must still be specified. We follow ref.\cite{matsen97} and
choose

\begin{eqnarray}
\Phi_0({\bf r})=&\frac{1}{2}[1-\cos(\pi x/\epsilon)],&\qquad 0\leq
x\leq\epsilon\nonumber \\
              =&1,\qquad\qquad &\epsilon\leq x\leq d_0-\epsilon
\nonumber \\
              =&\frac{1}{2}[1-\cos(\pi(d_0-x)/\epsilon)], \qquad
&d_0-\epsilon\leq x\leq d_0
\label{phizero}
\end{eqnarray}
and
\begin{eqnarray}
h({\bf r})=&\frac{\Lambda aN^{1/2}}{\epsilon}[1+\cos(\pi x/\epsilon)],
\qquad &0\leq x\leq \epsilon\nonumber \\
          =&0, \qquad\qquad &\epsilon\leq x\leq d_0-\epsilon,\nonumber \\
          =&\frac{\Lambda aN^{1/2}}{\epsilon}[1+\cos(\pi
(d_0-x)/\epsilon)],\qquad &d_0-\epsilon\leq x\leq d_0,
\label{surfacefield}
\end{eqnarray}
where $\epsilon$ is a small distance given below, and $\Lambda$ is the
strength of the surface field. A positive value of $\Lambda$ causes the
surface of the system to prefer the $A$ component.

Because of the three-dimensional nature of the problem, we have chosen
to solve the self-consistent equations in real-space, following the
pseudo-spectral method of Rasmussen and co-workers
\cite{rasmussen02,tzeremes02,rasmussen04}. We have taken a grid which is
$192\times 32\times 64$ in the $x,y,z$ directions respectively. The
length $\epsilon$ is chosen to be sufficiently
small to
ensure that it does not affect phase behavior. It is discussed further,
below. In examining the phase
of cylinders parallel to the plates, whose axes are in the $z$
direction,
we have investigated several arrangements of the cylinders, and varied
the distances between them, both in the $x$ and $y$ directions, to ensure
that we have found the minimum free energy configuration.

\section {RESULTS}

The parameter space of our system is large, as its state is specified by
five variables, the choice of which is not unique.  They are the temperature, or
equivalently, $\chi N$, the applied electric field, $E_{ext}$, the
chemical potential, $\mu$, or the film thickness $d_0$, the surface
field strength, $\Lambda N$, and the copolymer architecture, $f_A$.
In order to demonstrate the various morphologies, we have chosen to fix
$\chi N=18$, and the architecture at $f_A=0.29$ which corresponds to
the fraction of methyl methacrylate in the
poly-(methyl methacrylate)(PMMA)/polystyrene(PS) diblocks
used by Thurn-Albrect et al.
\cite{talbrecht00}. This leaves three parameters whose effects can be
explored. We have chosen to vary $E_{ext}$ and $d_0$, and have examined
two surface field strengths, $\Lambda
N=0.2$, and 0.5. To understand whether these surface fields are weak
or strong, the surface energy of $n$ polymers subject to these fields,
$U_{su}$, can be compared to the
electrostatic field energy of these polymers in volume $\Omega$.
The former of these energies is $U_{su}=nk_BTN\Lambda$, while
in the bulk cylindrical phase, the latter is
$U_{el}=\kappa_0\epsilon_0E_{ext}^2\Omega/2$, where
\begin{equation}
\kappa_0=\kappa_Af_A+\kappa_B(1-f_A),
\end{equation}
is the average dielectric constant of the system. Thus we consider the
ratio
\begin{eqnarray}
\frac{U_{el}}{U_{su}}&=&\frac{\epsilon_0\kappa_0E_{ext}^2\Omega}{2k_BT\Lambda
Nn},\nonumber\\
                    &=&\frac{\kappa_0 E_{ext}^2}{2\Lambda N{\cal
E}^2}\nonumber,\\
                    &=&\frac{\kappa_0{\hat E}_0^2}{2\Lambda N},
\end{eqnarray}
where we have introduced the convenient scale of electric field
\begin{equation}
{\cal E}\equiv\left(\frac{k_BT n}{\epsilon_0\Omega}\right)^{1/2},
\label{scale}
\end{equation}
and the value of the external field in these units,
${\hat E}_0\equiv E_{ext}/{\cal E}.$
At typical temperatures, $T\approx 430K$, and for a typical volume per
polymer chain of 100 nm$^3$, this unit of electric field is
${\cal E}\approx 82$V$/\mu$m. At experimental temperatures around
160$^\circ$C the dielectric constants appropriate to the PMMA-PS
copolymer, with PMMA being the $A$ block and PS the $B$ block
are\cite{talbrecht00,xu04,tsori03} $\kappa_A=6.0$ and $\kappa_B=2.5,$
from which $\kappa_0\approx 3.52$. Thus $U_{el}/U_{su}\approx 1.8{\hat
E}_0^2/{\Lambda N}$. At external fields of order of tens of volts per
micron, the surface field energies we consider are comparable
to the electrostatic energies, and are therefore smaller than the rather
large surface fields \cite{mansky97} of the experiments of Xu et al. \cite{xu05}.

We first consider the system with surface fields of $\Lambda N=0.5.$ We
have calculated, for a given thickness, $d_0$, the excess free energies
of various parallel, perpendicular, and intermediate phases. That with
the lowest excess free energy is the equilibrium one. Rather than plot
the excess free energy, $\sigma$, directly vs. applied field, we note
from eq. \ref{sigma} and the fact that the electrostatic contribution to the
bulk free energy of the
cylindrical phase is simply $-\epsilon_0\kappa_0E_{ext}^2/2,$
that for determining the globally stable phase,
we can equally well plot the dimensionless surface free energy
\begin{equation}
f_n({\hat E}_0)\equiv \frac{F_{tot}}{nk_BT}+\frac{1}{2}\kappa_0{\hat
E}_0^2.
\label{fsubn}
\end{equation}
The derivative of this function with respect to the
dimensionless external field
${\hat E}_0$ can be found from eq. \ref{displacement} to be
the negative of the dimensionless excess displacement field
\begin{eqnarray}
{\hat D}_s&\equiv& \frac{D_s}{d_0\epsilon_0{\cal E}},\nonumber \\
          &=&-\frac{\partial f_n({\hat E}_0)}{\partial {\hat E}_0}.
\end{eqnarray}

In fig 1. we show the excess free energy $f_n$ for several
phases as a function of the dimensionless external electric field for a
thickness $d_0=7.5N^{1/2}a.$
First we note, as anticipated, that the surface free
energy increases with external field, indicating that the surface
contribution to the displacement is, in fact, negative {\em i.e.} in the
opposite direction of the applied field. Secondly, we see that at this
thickness, the stable morphology at small external fields is one with
five layers of cylinders parallel to the plates. As the field is
increased, a first order transition occurs to a mixed state in which the
surface contribution to the displacement field is less negative.
At sufficiently large fields, there is a second transition to the state
in which all cylinders are parallel to the external field. The excess
displacement field is even less negative, but is not zero due to the
distortion of the cylinders near the plates. By repeating such
calculations as a function of thickness, we obtain the phase diagram
shown in fig. 2. Because the film thickness is a smooth
function of the chemical potential, this diagram reflects much of what
was anticipated earlier in our discussion of the phase diagram as a
function of electric field and chemical potential. Note that for films of
sufficient thickness to accommodate more than two layers of parallel
cylinders, the surface field is sufficiently strong to impose a surface
ordered layer which results in the existence of
an intermediate phase.  At higher
fields, this surface ordered layer becomes too costly, and disappears
resulting in the perpendicular phase.  As anticipated,
the electric field which brings about this latter transition varies relatively
weakly with film thickness. For sufficiently thin films, both
orientations of cylinders, characteristic of the intermediate phase,
cannot be maintained leading to the elimination of the intermediate phase.

In determining the phase boundary between the perpendicular and
intermediate phases, and between the perpendicular and the cylindrical
phase with two parallel layers, we have set the characteristic
length, $\epsilon$, of the surface field to a constant value
$3N^{1/2}a/16.$
It is not difficult to see from eqs. \ref{phizero}, \ref{surfacefield} that
if $\epsilon$ is to equal an integer number $n$ of grid points, of
which there are 192 in the $x$ direction, then
the thickness $d_0/N^{1/2}a$ must take the discrete values $36/n$.
These discrete points are
shown in fig. 2 where they are connected by a spline fit, shown dotted, to
guide the eye. In contrast to the relatively smooth variation with
thickness of the
electric field at the transition from the perpendicular phase, that at
the transition from the parallel phase varies rapidly with thickness.
Hence the discrete values of thickness imposed by a constant
$\epsilon$ is restrictive. Further, the behavior of this critical field
with thickness has already been  anticipated from our discussion
of the thermodynamics of the transition. For these transitions,
therefore, we chose $\epsilon$ to be a constant fraction of film
thickness, $\epsilon/d_0=1/32.$ It is readily seen, again from
eqs. \ref{phizero}, \ref{surfacefield}, that this choice permits the film
thickness $d_0/N^{1/2}a$ to vary continuously. We have compared the two
values of the critical fields obtained from these two different choices
of $\epsilon$ at the discrete values of $d_0=36/n$
permitted by a constant $\epsilon$, and found essentially no  difference.

We next present density profiles of the various phases
for a distance between plates of $d_0=7.5aN^{1/2}$. When the applied
field is zero, the stable phase is that shown in fig. 3 which we denote
as having five parallel layers.
The cut is in the $x$ (vertical), $y$ (horizontal) plane, a cut normal to
the plane of the plates. The gray scale shows the volume fraction
of $A$ monomer from
0.0 (lightest) to 1.0 (darkest) in four bins of width 0.25. The $A$ block
is favored by the surface field of strength $\Lambda N=0.5$. The
distortion of the cylinders at the surface is clear. A cut in the $yz$
plane, parallel to the plates, (not shown), confirms that these are,
indeed cylinders with axis along the $z$ direction.

With an applied
field of sufficient strength, the intermediate state becomes globally
stable. A cut in the $xy$ plane is shown
in fig. 4 for a value ${\hat E}_0=0.49$. Most of the film
thickness is occupied by cylinders which are perpendicular to the
plates, and a cut in the $yz$ plane halfway between the plates, (not
shown), confirms that the cylinders are in a hexagonal array. It might
be expected from fig. 4 that, adjacent to the surface,
one would find a distorted layer of cylinders, just as in
fig. 3. That this is {\em not} the case is shown by a cut
through the intermediate phase in a $yz$ plane very near the surface,
fig. 5. One sees instead that there is an {\em hexagonal}
array, such that the surface-favored $A$ monomers occupy most of the
surface layer. Clearly the hexagonal arrangement on the surface derives
from the hexagonal, perpendicular, arrangement of cylinders in most of
the film. Comparison of
a sequence of $yz$ cuts confirm that the symmetry axes of the upright
cylinders in this intermediate phase coincide with the symmetry axes of
the hexagonal array at the surface; i.e. the $A$-rich cores of the
cylinders stand directly above the $B$-rich centers of the surface
pattern. That the $A$ monomers near the surface are located
{\em outside} of the cylinders thereby occupying a larger area in
the unit cell is clearly an effect due to the surface
field. Because the $A$ monomers are the minority component, $f_A=0.29$,
this configuration cannot extend too far into the film.

For larger values of the electric field, the perpendicular phase is
attained. Fig. 6 shows the density profile in the $xy$ plane
at a field of ${\hat
E}_0=0.78.$ One clearly sees the distortion of the cylinders produced by
the surface.

We next show, in fig. 7, the phase diagram for the weaker surface field
of strength
$\Lambda N=0.2.$  Again, we chose $\epsilon=3N^{1/2}a/16.$
One sees that the weaker surface field can no longer impose a surface
morphology which differs from, yet coexists with, that of the remainder
of the film, even for ones sufficiently
thick to accommodate seven parallel layers.  Thus there is no
intermediate phase. Instead, with increasing external field,
transitions from all cylindrical phases shown, up to seven
parallel layers, are directly to the perpendicular phase.
The maximum critical fields
of the transition between parallel and perpendicular phases
vary as $1/{\sqrt d_0},$ as expected. This behavior is shown in the inset.
The six maximum critical fields are shown by the open triangles
which fall on a straight line in this plot.

\section{Discussion}

We have applied self-consistent field theory to a planar film of
cylindrical-forming block copolymers. The plates between which the film
is adsorbed prefer the minority $A$ component.
In the absence of an electric field, this causes the morphology
to be that of cylinders
whose axes are parallel to that of the plates. We have studied the
equilibrium phase diagram in the presence of an applied electric field
which tends to align the cylinders perpendicular to the plates,
and have solved the Maxwell equations without approximation.
The behavior as a function of field and film thickness is,
for the most part, similar to
that of lamellar-forming dibocks. As argued above, this is because the
underlying physics, that of the competition between bulk and surface
phases, is the same.  In particular for weak surface
fields or thin
films, the surface field cannot impose a surface morphology which
differs from the rest of the film, so that there is no intermediate
phase. The first-order transition with increasing electric field
from the phase of parallel cylinders to
that of perpendicular cylinders is direct.
However for stronger surface fields and thicker films, an intermediate
phase does appear, one which is characterized by a boundary layer near the
plates.

The nature of this boundary layer is very different for
cylinder-forming diblocks than for lamellar-forming ones for which this
layer is simply a modulated lamellae.
We considered different
possibilities for this boundary layer. Because of the experimental
report of a signal indicating the presence of both
perpendicular and parallel cylinders\cite{talbrecht00},
we considered whether a layer near the
plates might consist of parallel cylinders, but found that it did
not. In retrospect, this
is not difficult to
understand. First, the perpendicular
cylinders aligned with the field could not
connect in a simple periodic manner with cylinders aligned parallel to the
plates as the two arrays would be incommensurate. Thus one of the two
arrays would have to be distorted, with a concomitant
increase of free energy. Second, the
substrate does not favor parallel cylinders {\em per se}, rather it favors
one monomer over the other, the minority $A$ monomer in our case. In zero
electric field, parallel cylinders are favored over perpendicular ones
because the area of $A$ monomers presented to the plates is larger in the
former than in the latter. The ratio of areas is
${\cal A}_{par}/{\cal A}_{perp}\approx\ 1/f_A^{1/2}>1$.
In a non-zero field which favors
the intermediate phase, however, the area presented to the plates by
parallel cylinders is less than that presented in the configuration of
hexagonal symmetry which
we obtained and which is shown in Fig. 5. In this case the ratio of areas is
${\cal A}_{par}/{\cal A}_{int}\approx f_A^{1/2}/(1-f_A)<1$.
We also note that this particular pattern of the surface layer derives
from the fact that the minority $A$ component is favored by the
surface. Were the $B$ component favored, as is the case in the system of
Xu et al. \cite {xu05}, one would expect the minority
core of the cylinders to extend all the way to the plates over a range
of surface fields, something which could be technologically useful. Only
for very strong fields would the plates be essentially covered by the
majority $B$ component which would require the interior of the
cylinders containing the $A$ component to be truncated near the surface.
This
observation is in agreement with the dynamic self-consistent field
theory results as seen in Fig. 11(b) of Xu et al. \cite{xu05} and Fig. 5
of Lyakhova et al. \cite{lyakhova06}.

We have found a transition from the intermediate phase to the
perpendicular phase for which the electric field needed to bring about
the transition varies only weakly with film thickness. It would seem
that this could be identified with the transition observed by
Thurn-Albrecht et al\cite{talbrecht00}. There are, however, two
difficulties with this interpretation. The first is that the electric
field values at the transition are, in our calculation, larger than
those observed in experiment. Further, the strength of the surface field
in our calculation is weaker than that estimated in experiment. Were we
to calculate critical electric fields for
the stronger substrate fields of experiment, we would obtain even larger
values as the critical electric field is expected to
increase as the square root of
the surface field. This comparison is reminiscent of that between calculated
\cite{lin05,matsen05} and
experimental \cite{xu04} electric field values for the transition from a
body-centered cubic phase to a cylindrical phase in which the former
were also much larger than the latter. Explanations for this have been
proposed \cite{tsori03,matsen06} which may resolve this
difficulty. Even so, there remains the fact that, in
the calculated intermediate phase, there are no cylinders in the parallel
orientation. Therefore one would not observe {\em in equilibrium} the existence of
both parallel and perpendicular cylinders along the phase coexistence
of the intermediate and perpendicular phases.
The resolution of this difficulty
may be that, as the system cools from
the disordered into
the ordered phase, cylinders are nucleated near the plates in a
configuration parallel to them while the major part of the film
nucleates cylinders aligned with the electric field. Rearrangement of cylinders
near the plates into that boundary layer which we predict to be the
equilibrium one would presumably be a very slow process. If the boundary
layer did not attain its equilibrium configuration, the experimental
signal which detects both orientations of cylinders could be understood.

We thank K. \O. Rasmussen for helpful correspondence, and D. A. Andelman
for seminal conversations. This work was
supported by the U.S.-Israel Binational Science Foundation
(B.S.F.) under grant No. 287/02 and the
National Science Foundation under grant No. DMR-0503752.

\clearpage


\clearpage
\section{Figure Captions}

Fig. 1 \ Dimensionless excess surface free energy, $f_n$,
eq. \ref{fsubn}, as a function of applied field, ${\hat E}_0$, for
thickness $d_0=7.5N^{1/2}a$. The value of $\chi
N=18$, and the surface field is $\Lambda N=0.5$. The free energy of the
perpendicular phase is shown with  a solid line, that of the intermediate
phase with a dashed-dotted line, and that of the parallel phase of five
layers is shown with a dotted line.

Fig. 2 \ Phase diagram as a function of electric field, ${\hat E}_0$,
and thickness, $d_0/N^{1/2}a$. The value of $\chi N=18$,
and the surface field is $\Lambda N=0.5$. Parallel phases are denoted by
a roman numeral corresponding to the number of cylinders in the
film. The intermediate and perpendicular phases are marked ``Intermed'',
and ``Perp'' respectively. Discrete calculated points on the boundary between
perpendicular and intermediate phases are shown by solid dots.
The dotted line is a spline fit to them.

Fig. 3\ Density profile in the $xy$ plane of the parallel phase with
five layers of cylinders. The thickness is $d_0/N^{1/2}a=7.5$, and the
applied field is zero. The gray scale shows the volume fraction
of $A$ monomer from
0.0 (lightest)to 1.0 (darkest) in four bins of width 0.25.

Fig. 4\ Density profile in the $xy$ plane of the intermediate phase. The
thickness is $d_0/N^{1/2}a=7.5$ and the applied field is now
${\hat E}_0=0.49$

Fig. 5\ Density profile of the intermediate phase
shown here in a $yz$ plane very close to the plates.

Fig. 6\ Density profile of the perpendicular phase in the $xy$ plane.  The
thickness is $d_0/N^{1/2}a=7.5$ and the applied field is now
${\hat E}_0=0.78$

Fig. 7 Phase diagram as a function of electric field, ${\hat E}_0$ and
thickness $d_0/N^{1/2}a$ for a surface field $\Lambda N=0.2$ and $\chi
N=18$. Inset shows as open triangles
six calculated maximum electric field values at the
transition from parallel to perpendicular phases plotted vs. $1/d_0^{1/2}$.
\clearpage
\begin{figure}[!t]
\includegraphics[scale=1.0,bb=45 45 410 302,clip]{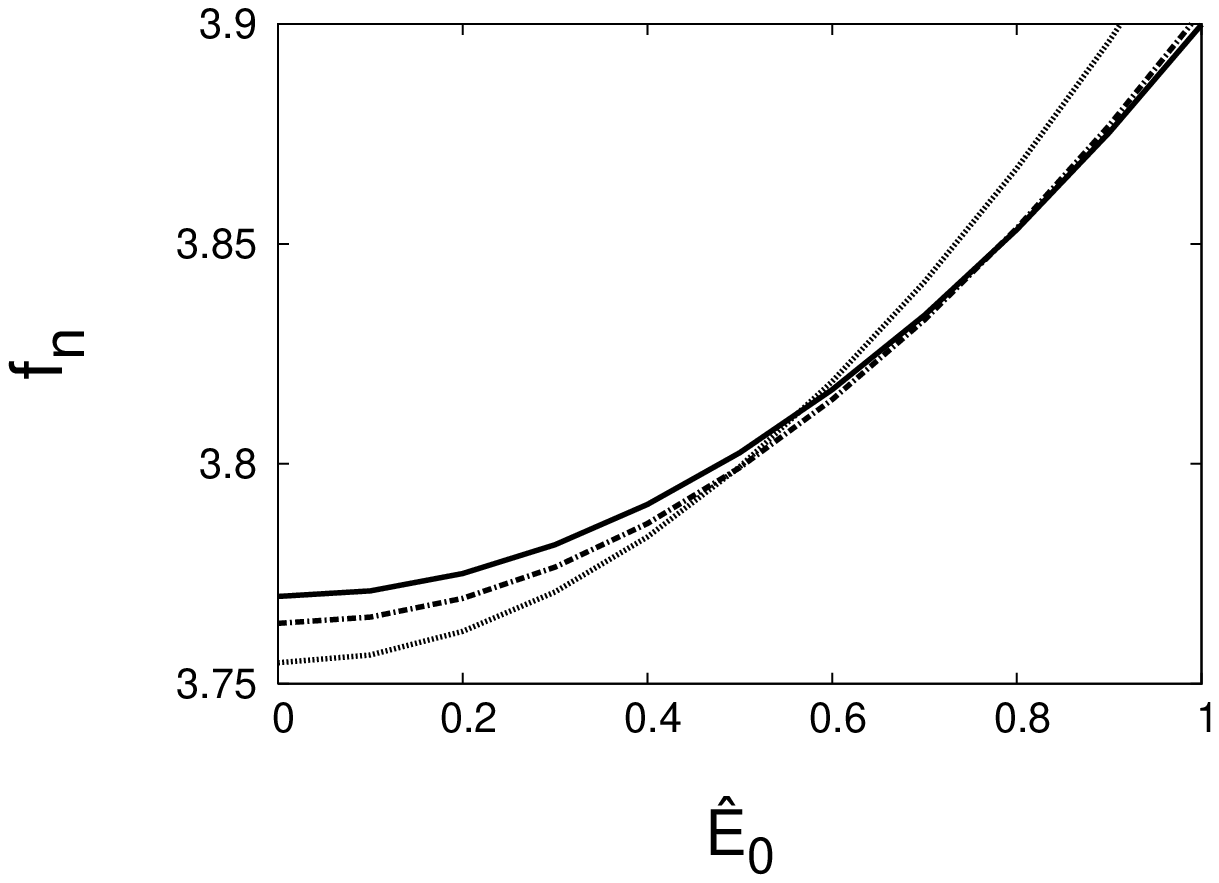}
\label{fig1}
\end{figure}

\begin{center}
\textrm{
\newline
\newline
\newline
\newline
\newline
\newline
\newline
Chin-Yet Lin and M. Schick, Fig. 1 }
\end{center}
\clearpage

\begin{figure}[!t]
\includegraphics[scale=1.0,bb=45 45 410 302,clip]{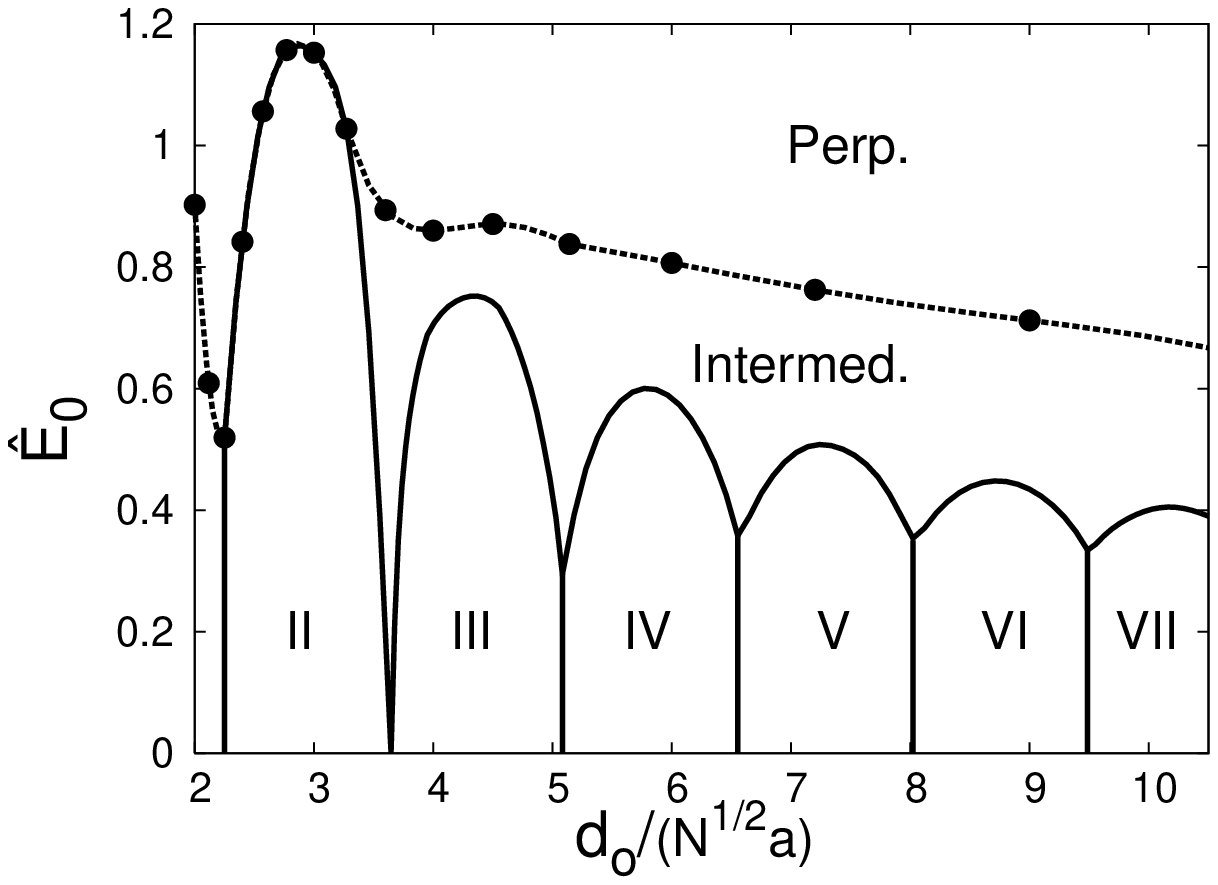}
\label{fig2}
\end{figure}

\begin{center}
\textrm{
\newline
\newline
\newline
\newline
\newline
\newline
\newline
Chin-Yet Lin and M. Schick, Fig. 2 }
\end{center}
\clearpage

\begin{figure}[!t]
\includegraphics[scale=1.0,bb=0 0 107 383,clip]{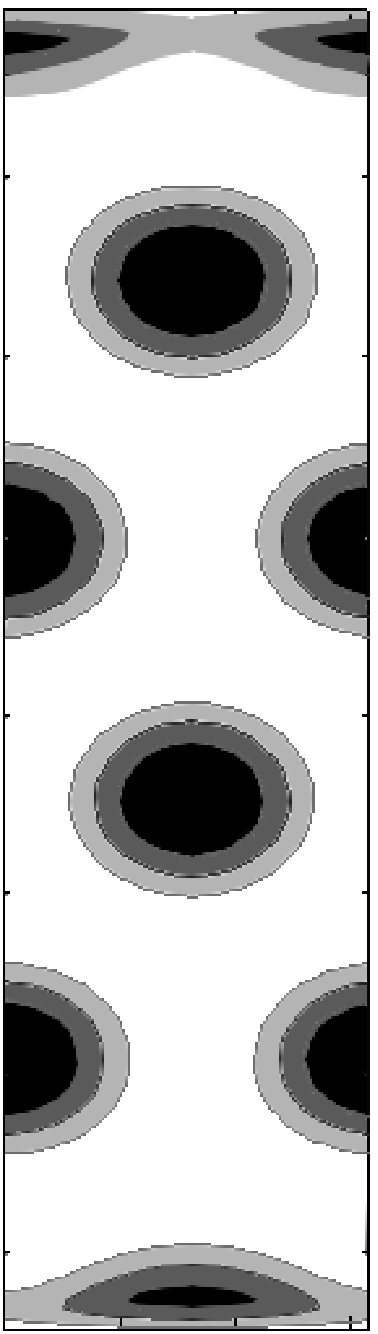}
\label{fig3}
\end{figure}

\begin{center}
\textrm{
\newline
\newline
\newline
\newline
\newline
\newline
\newline
Chin-Yet Lin and M. Schick, Fig. 3 }
\end{center}
\clearpage
\begin{figure}[!t]
\includegraphics[scale=1.0,bb=0 0 107 383,clip]{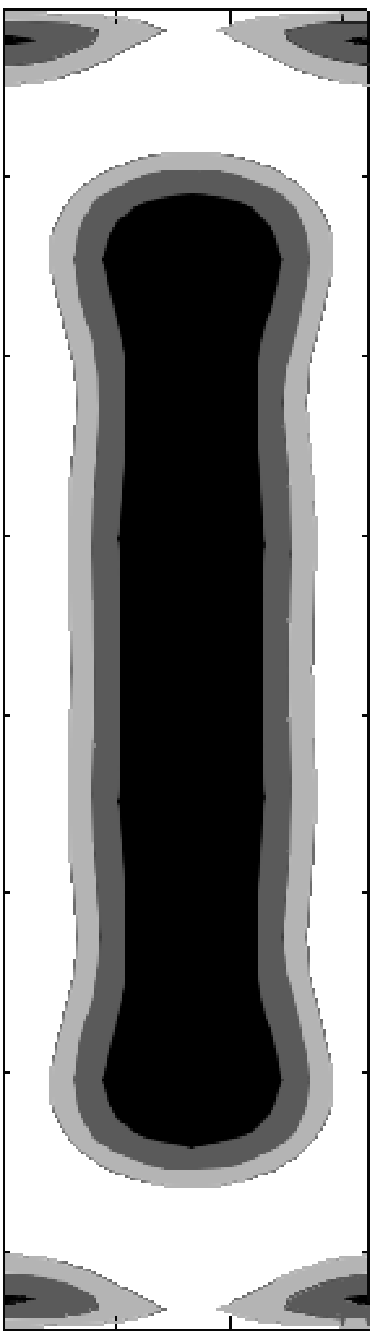}
\label{fig4}
\end{figure}

\begin{center}
\textrm{
\newline
\newline
\newline
\newline
\newline
\newline
\newline
Chin-Yet Lin and M. Schick, Fig. 4 }
\end{center}

\clearpage
\begin{figure}[!t]
\includegraphics[scale=1.0,bb=0 0 383 218,clip]{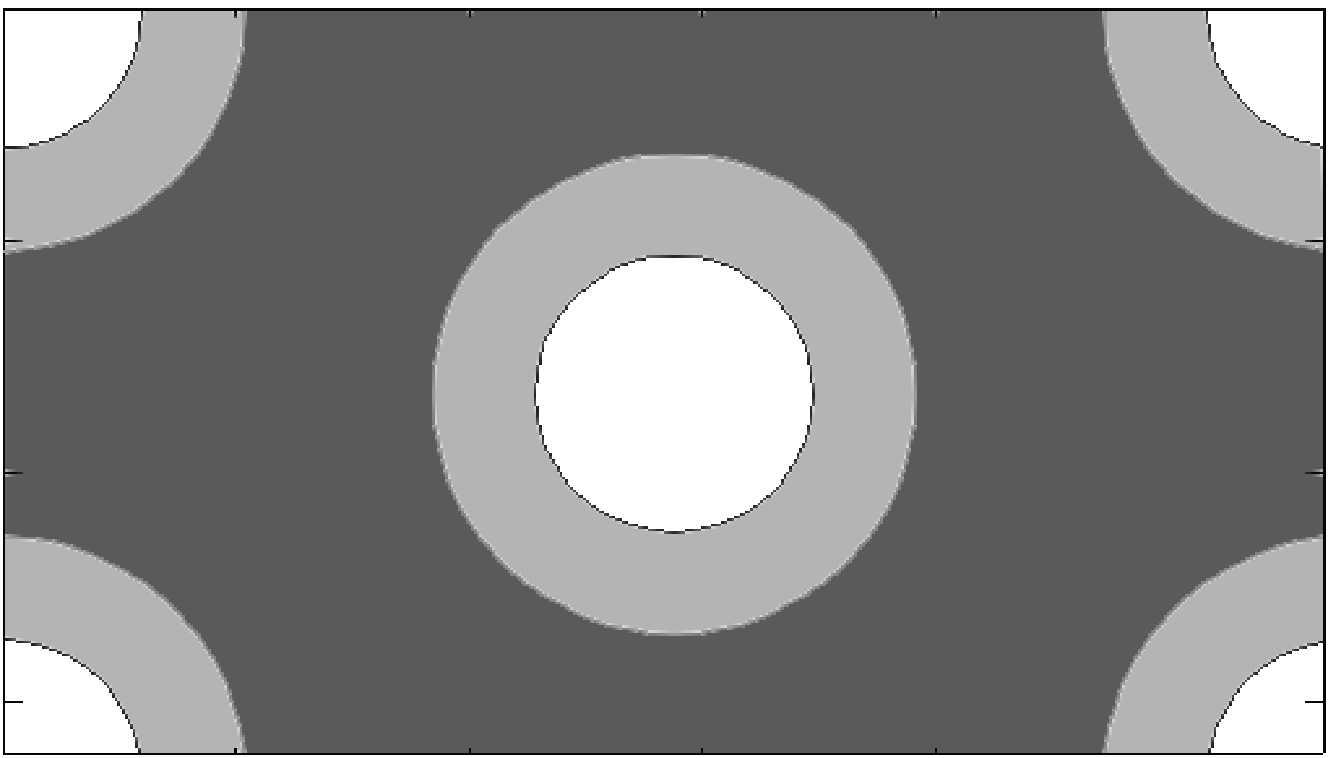}
\label{fig5}
\end{figure}
\begin{center}
\textrm{
\newline
\newline
\newline
\newline
\newline
\newline
\newline
Chin-Yet Lin and M. Schick, Fig. 5 }
\end{center}

\clearpage
\begin{figure}[!t]
\includegraphics[scale=1.0,bb=0 0 107 383,clip]{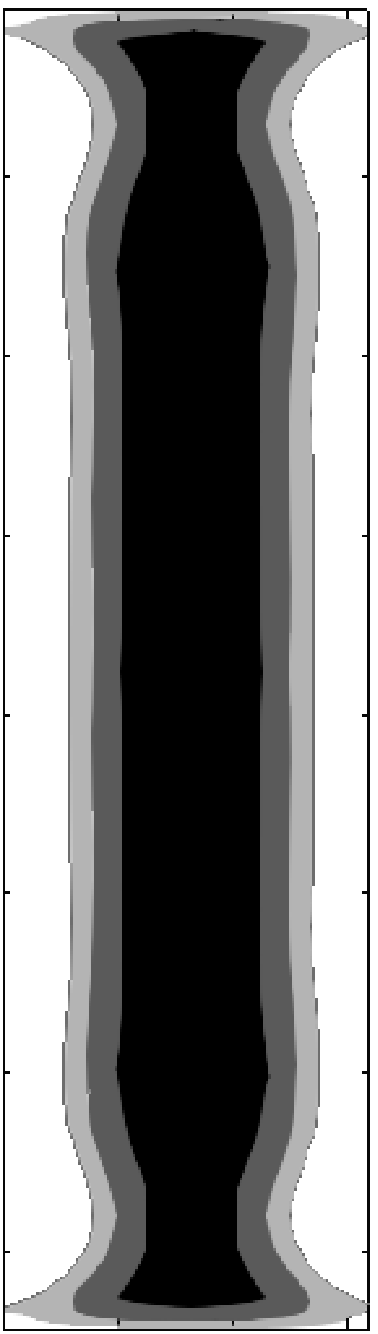}
\label{fig6}
\end{figure}
\begin{center}
\textrm{
\newline
\newline
\newline
\newline
\newline
\newline
\newline
Chin-Yet Lin and M. Schick, Fig. 6 }
\end{center}

\begin{figure}[!t]
\includegraphics[scale=1.0,bb=45 45 410 302,clip]{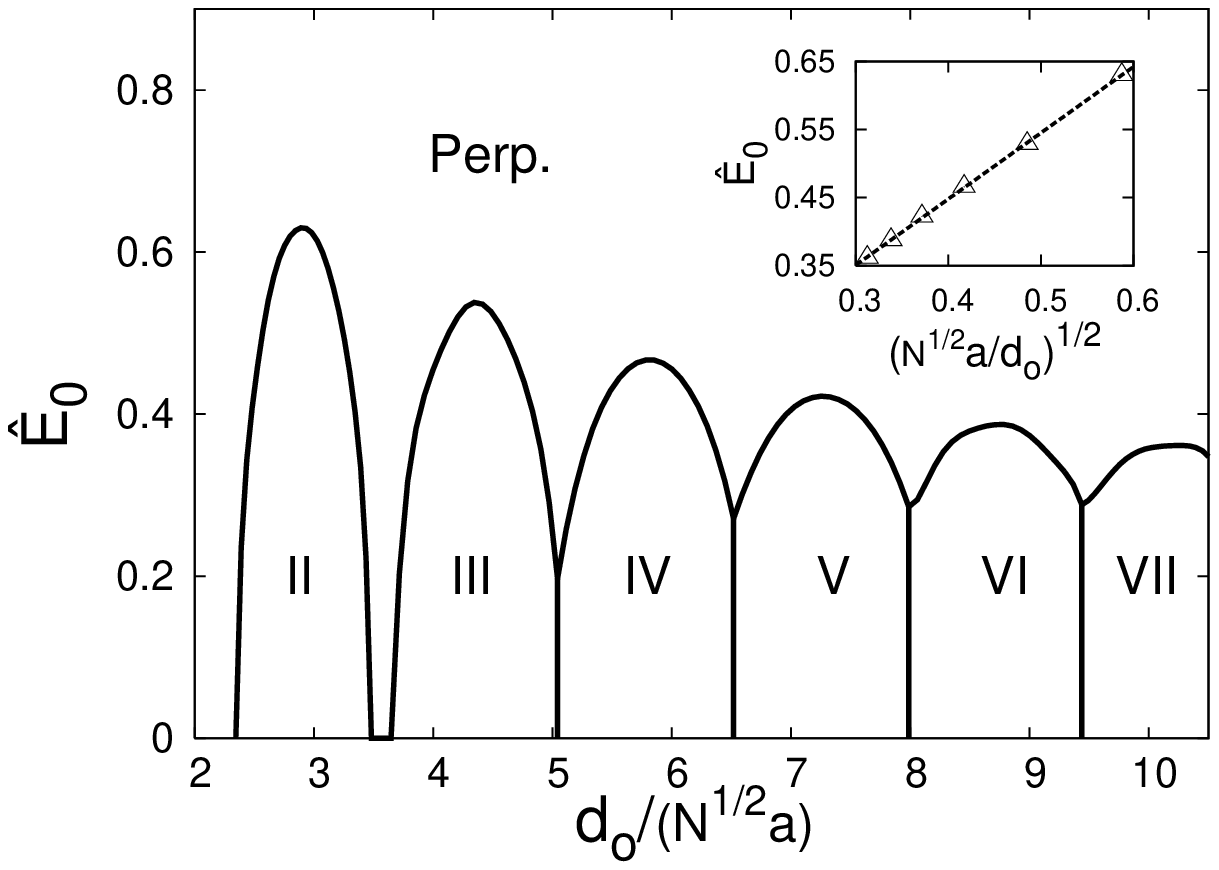}
\label{fig7}
\end{figure}

\begin{center}
\textrm{
\newline
\newline
\newline
\newline
\newline
\newline
\newline
Chin-Yet Lin and M. Schick, Fig. 7 }
\end{center}
\clearpage
\clearpage

\end{document}